\documentstyle[aps,epsf,prl]{revtex}

\begin{document}
\draft
\title{A New Iterative Perturbation Scheme for Lattice Models with 
Arbitrary Filling}
\author{Henrik Kajueter and Gabriel Kotliar}
\address{Department of Physics, Rutgers University, Piscataway, 
NJ 08855-0849, USA}
\date{\today}
\maketitle
\begin{abstract}
We derive a new perturbation scheme for treating the large d limit of
lattice models at arbitrary filling. The results are compared with 
exact diagonalization data for the Hubbard model and found to be in 
good agreement.
\end{abstract}
\narrowtext
\twocolumn
\section{INTRODUCTION}
In recent years there has been a renewed interest in the study of
doped transition metal oxides like $La_{1-x}Ti_{x}O_3$. These
materials exhibit interesting 
phenenoma like the correlation induced metal insulator
transition.
Although there are several experimental data available 
right now \cite{tokura,tokura1,tokura3}
it is still quite difficult to tackle
these substances theoretically.
Realistic
 models have to take into
account several bands and are to be explored at finite doping.

The most promising way towards a theoretical description,
perhaps, is the limit of large spatial
dimensions \cite{cumetzner}, which defines a dynamical mean field
theory for the problem.  This limit can be mapped onto an impurity
model together with a selfconsistency condition which is
characteristic for the specific model under consideration \cite{cugeorges}.
The mapping allows to apply several numerical and analytical
techniques which have been developed to analyse impurity models over
the years. There are different approaches which have been used for
this purpose: qualitative analysis of the mean field 
equations \cite{cugeorges},
quantum Montecarlo methods \cite{cujarell1,curzk,cugkrauth},
 iterative perturbation theory (IPT) \cite{cugeorges,zrk},
 exact diagonalization methods \cite{c4}, and the projective self-consistent
method, a renormalization techique \cite{moeller}.
However, each of these methods has its shortfalls. While quantum
Montecarlo calculations are not applicable in the zero temperature
limit, the exact diagonalization methods and the the projective 
self consistent method yield only a discrete number of pols for the
density of states. Moreover, 
the  computational requirements of the
exact diagonalization and the quantum Montecarlo
methods   are such that they can only be implemented for
the simplest hamiltonians. To carry out realistic calculations
it is necessary to have an accurate but fast algorithm for
solving the Impurity model. 
In this context
iterative perturbation theory has turned out
to be a useful and reliable tool for the case of half filling \cite{long,v2o3}.
However, for finite doping the naive extension of the IPT scheme is
known to give unphysical results. 
There is still no method which can be applied away from half
filling and which at the same time is powerful enough to treat more
complicated models envolving several bands.

The aim of this paper is to close this gap by introducing a new
iterative perturbation scheme which is applicable at arbitrary
filling. For simplicity, we treat the single band case here. But we 
believe that the ideas can be generalized to more complicated models
involving several bands.

The (asymmetric) Anderson impurity model
\begin{eqnarray}\label{eu1}
H_{imp}&=&\epsilon_f\,\sum_\sigma f_\sigma^+f_\sigma\,+\,
\sum_{k,\sigma} \epsilon_k\, c_{k\sigma}^+ c_{k\sigma} \nonumber \\
&+& \sum_k V_k
\left(c_{k\sigma}^+f_{\sigma}+f_{\sigma}^+c_{k\sigma}\right)\,
+\,U f_\downarrow^+ f_\downarrow\,
 f_\uparrow^+ f_\uparrow
\end{eqnarray}
describes an impurity $(f_\sigma)$ coupled to a bath of conduction
electrons $(c_{k\sigma})$. The hybridization function is given by 
$\Delta(\omega):=\sum_k\frac{V_k^2}{\omega-\epsilon_k}$.
Once a solution is known for arbitrary parameters a 
large number of lattice models can be solved by  iteration.
An example is the 
Hubbard hamiltonian:
\begin{equation}\label{eu2}
H=-\frac{t}{\sqrt{z}}\, \sum_{\langle ij \rangle,\sigma} c_{i\sigma}^+
c_{j\sigma}\ + \ U\sum_{i}
n_{i\uparrow}\, n_{i\downarrow}
\end{equation}
which can  actually serve as an  effective hamiltonian for the
description of doped transition metal oxides \cite{kkm}.
On a Bethe lattice with infinite coordination number $z$
the Hubbard model  is connected to the impurity
model  by the following selfconsistency condition:
\begin{equation} \label{ue20}
\Delta(\omega)=t^2\, G(\omega) 
\end {equation}
and $\epsilon_f=-\mu$. The mapping requires that the propagator of the
lattice problem is given by the impurity Green function
($G=G_f$\/). Below we set $D:=2t=1$.

In the next section we will derive the perturbation scheme for the
impurity model. Afterwards  the scheme is applied
to the Hubbard model (section \ref{su3}). 
Some results for the doped system are presented
and the accuracy of our scheme is discussed.
We conclude with a summary and an outlook on further extensions
(section \ref{su4}).

\section{Derivation of the approximation scheme for the single
impurity model}\label{su2}
In this section, we derive the approximation scheme
which, given the hybridization function 
$\Delta(\omega)$ and the impurity 
level $\epsilon_f$, provides a solution of model (\ref{eu1}).
For simplicity, we assume that there is no magnetic symmetry breaking
($n_{\sigma}=n_{-\sigma}=n$\/). We also restrict us to zero temperature.
The procedure  is an
extension of the ordinary IPT scheme to finite doping. The success of
IPT at half filling can be explained by that it becomes exact not only
in the weak but also in the strong coupling limit \cite{long}.
Moreover, this approach captures the right low and high frequency
behavior so that we are dealing with an interpolation scheme between
correct limits.

The idea of our approach is to construct a self energy expression which
retains these features at arbitrary doping and reduces at half filling
to the ordinary IPT result.

Ordinary IPT approximates the self energy by its  second order
contribution: $\Sigma(\omega) \approx Un
+\tilde{\Sigma}_0^{(2)}(\omega)$ where
\begin{eqnarray}\label{eu3}\nonumber
\tilde{\Sigma}^{(2)}_0(\omega)&:=& U^2
\int_{-\infty}^{0}d\epsilon_1\,\int_{0}^{\infty}d\epsilon_2d\epsilon_3\,
\frac{\rho^{(0)}(\epsilon_1)\,\rho^{(0)}(\epsilon_2)\,
\rho^{(0)}(\epsilon_3)}{\omega+\epsilon_1-\epsilon_2-\epsilon_3-i\eta}
\nonumber \\
&+& U^2
\int_{0}^{\infty}d\epsilon_1\,\int_{-\infty}^{0}d\epsilon_2d\epsilon_3\,
\frac{\rho^{(0)}(\epsilon_1)\,\rho^{(0)}(\epsilon_2)\,
\rho^{(0)}(\epsilon_3)}{\omega+\epsilon_1-\epsilon_2-\epsilon_3-i\eta}
\end{eqnarray}
with $\rho^{(0)}=\frac{1}{\pi}\mbox{Im} G_0$.
 Here, the (advanced) Green
function
$ G_0(\omega)$ is defined by
\begin{equation}\label{eu4}
  G_0(\omega):=\frac{1}{\omega+\tilde{\mu}_0-\Delta(\omega)}
\end{equation}
The parameter $\tilde{\mu}_0$ is given by
$-\epsilon_f-Un$. In particular it vanishes at half filling.
The full Green function follows from
\begin{equation}\label{eu5}
G_f(\omega)=\frac{1}{G_0^{-1}-\tilde{\mu}_0-\epsilon_f-\Sigma(\omega)}
\end{equation}

To ensure the correctness of this approximation scheme in different
limits, we modify the self energy functional 
as well as the definition of the parameter $\tilde{\mu}_0$.

We start with an ansatz for the self energy:
\begin{equation}\label{eu6}
\Sigma_{int}(\omega)=Un\,+\,\frac{A\,\tilde{\Sigma}_0^{(2)}(\omega)}{1-
B\,\tilde{\Sigma}_0^{(2)}(\omega)}
\end{equation}
Here, $\tilde{\Sigma}_0^{(2)}(\omega)$ is the normal second order
contribution defined in equation (\ref{eu3}). We determine the
parameter $A$ from the condition that the self energy has the exact
behavior at high frequencies. Afterwards, $B$ is determined from 
the atomic limit.

The leading behavior for large $\omega$ can be obtained by expanding
the Green function into a continuous fraction \cite{gordon}:
$G_f(k,\omega)=1/(\omega-\epsilon_f-\alpha_1-
\frac{\alpha_2-\alpha_1^2}{\omega+\ldots})$.
Here, $\alpha_i$ marks the $i$th order moment of the density of states.
One can compute these quantities by evaluating a commutator (see
\cite{nolting}). We obtain for our model 
$G_f(k=k_{fermi},\omega)
=1/(\omega-\epsilon_f-Un-\frac{U^2\,n(1-n)}{\omega+\ldots})$.
The leading term of the self energy is therefore given by
\begin{equation}\label{eu7}
\Sigma(\omega)=Un+U^2\,n(1-n)\ \frac{1}{\omega}+
O\left(\left(\frac{1}{\omega}\right)^2\right)
\end{equation}
Here, $n$ is the physical particle number given by 
$n=\int_{-\infty}^{0} d\omega\,\mbox{Im} G_f(\omega)$.
(\ref{eu7}) has to be compared with the large frequency limit of (\ref{eu3}):
\begin{equation}\label{eu8}
\tilde{\Sigma}^{(2)}_{0}(\omega)=U^2\,n_0(1-n_0)\ \frac{1}{\omega}+
O\left(\left(\frac{1}{\omega}\right)^2\right)
\end{equation}
where $n_0$ is a fictitious particle number determined from $G_0$
(i.~e.~$n_0=\int_{-\infty}^0 d\omega\,\mbox{Im} G_0(\omega)$\/).
From (\ref{eu6}), (\ref{eu7}), and (\ref{eu8}) we conclude
\begin{equation}\label{eu9}
A=\frac{n(1-n)}{n_0(1-n_0)}
\end{equation}
Chosing $A$ in this way guaranties that our self energy is correct to
order $\frac{1}{\omega}$. It should be noted from the continuous
fraction considered above that consequently the moments of the density
of states up to second order are reproduced exactly.

Next, we have to fix $B$. The exact  impurity Green
function for $V_k\to 0$ is given by \cite{brenig}
\begin{equation}\label{eu10}
G_f(\omega)=\frac{n}{\omega-\epsilon_f-U-i\eta}+\frac{1-n}{\omega-\epsilon_f-i\eta}
\end{equation}
This can be written as
$G_f(\omega)=1/(\omega-\epsilon_f-\Sigma_{atomic}(\omega))$
where
\begin{equation}\label{eu11}
\Sigma_{atomic}(\omega)=Un+\frac{n(1-n)\,U^2}{\omega-\epsilon_f-(1-n)U-i\eta}
\end{equation}
This expression is to be compared with the atomic limit of our ansatz
(\ref{eu6}). Since $\tilde{\Sigma}^{(2)}_0(\omega) \to
\frac{U^2\,n_0(1-n_0)}{\omega+\tilde{\mu}_0-i\eta}$, we obtain
\begin{equation}
B=\frac{(1-n)\,U+\epsilon_f+\tilde{\mu}_0}{n_0(1-n_0)\,U^2}
\end{equation}

Thus, the final result for our interpolating self energy is 
\begin{equation} \label{eu12}
\Sigma_{int}(\omega)= Un+\frac{ \frac{n\,(1-n)}{n_0\,(1-n_0)}\, 
   \tilde{\Sigma}^{(2)}_0(\omega)}{1-\frac{(1-n)\,U+\epsilon_f+
         \tilde{\mu}_0}{n_0\,(1-n_0)\,U^2}\, \tilde{\Sigma}^{(2)}_0(\omega)}
\end{equation}

Yet, $\tilde{\mu}_0$ is still a free parameter. We fix it imposing
the Friedel sum rule \cite{friedel}:
\begin{eqnarray} \label{eu13}
  n&=&\frac{1}{2} -\frac{1}{\pi}\mbox{arctg} 
    \left| \frac{-\epsilon_f-\Sigma_{int}(0)-
\mbox{Re}\Delta(0)}{\mbox{Im}\Delta(0)}\right|\nonumber\\
& &      +\int_{-i\infty}^{+i\infty}\frac{d\omega}{2\pi i}\ 
           e^{i\omega \, 0+} 
         G_f(\omega) \frac{\partial\Delta(\omega) }{\partial \omega}
\end{eqnarray}
This statement, which is equivalent to the Luttinger theorem \cite{luttinger}
$\int_{-i\infty}^{+i\infty} \frac{d\omega}{2\pi i} G_f(\omega)
\frac{\partial \Sigma_{int}(\omega)}{\partial \omega}=0$,
should be viewed as a condition on the zero frequency value of the
self energy to obtain the correct low energy behavior. 
The use of the Friedel sum rule is the main difference to
an earlier approximation scheme \cite{rodero} and is
essential to obtain a good
agreement with the exact diagonalization method.

So far, we considered three different limits: strong coupling, zero
frequency and large frequency. It remains to check the weak coupling
limit. Taking into account that $n=n_0$ and
$\tilde{\mu}_0=-\epsilon_f$ 
for $U=0$,
it follows that (\ref{eu12}) is indeed exact to order $U^2$.

The actual solution of the impurity model is determined by a pair 
($\tilde{\mu}_0$, $n$\/) which satisfies equations (\ref{eu3}), (\ref{eu4}),
(\ref{eu5}), (\ref{eu12}), and (\ref{eu13}).
For the numerical implementation Broyden's method
\cite{numerical_recipes},
 a generalization of
Newton's method, has turned out to be very powerful.
Defining two functions
$f_1(n,\tilde{\mu}_0):=n-\int_{-\infty}^{0} d\omega \mbox{Im}
G_f[n,\tilde{\mu}_0](\omega)$
and
$f_2(n,\tilde{\mu}_0):=n-n_{Friedel}[n,\tilde{\mu}_0]$
the impurity problem can be solved by 
searching for the zeros of $f_1$ and $f_2$
($n_{Friedel}$ is the particle number determined from the Friedel sum
rule).
The algorithm is very efficient as in most cases a
solution is found within 4 to 10 iterations.

\section{Application to the Hubbard model}\label{su3}
After treating  the
Anderson impurity model, we now apply the perturbation scheme to the
solution of the 
Hubbard model.
Starting with a guess for $\Delta(\omega)$ one can solve the impurity
model using the scheme described above. This yields a
 propagator $G_f=G$, which can be used to determine a new
hybridization function $\Delta(\omega)$ according to (\ref{ue20}).
The iteration is continued until convergence is attained.
It is most accurate to perform the calculation first on the
imaginary axis. Once the constants $A$ and $B$ in the interpolating self
energy are determined in this way, they can be used to perform the
iteration on the real axis. 

In the case of the Hubbard model, the Luttinger theorem takes the
simple form \cite{hartmann}
\begin{equation}\label{ue21}
    \mu_0=\mu-\Sigma_{int}(\omega=0),\mbox{\ \ \ } \mu_0:=\mu|_{U=0}
\end{equation}
This can be used to simplify the selfconsistency procedure if $\tilde{\mu}_0$
rather than $\mu$ is fixed. Starting with a guess for $G$ and $\mu$,
one can compute $G_0$, $n$, and $n_0$. Afterwards (\ref{eu12}) yields
$\Sigma_{int}(\omega)$ and a new $\mu$ is obtained from (\ref{ue21}).

\begin{figure}
\centerline{\epsfxsize=2.8truein
\epsffile{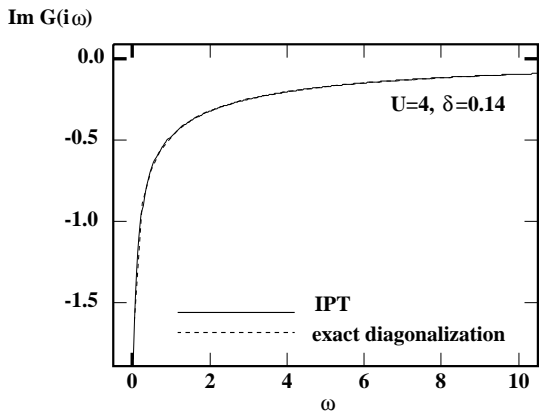}}
\caption{$\mbox{Im} G(i\omega)$ at $T=0$
for $U=4D$ and hole doping $\delta=0.14$: iterative perturbation scheme
(full line) vs.~exact diagonalization (dashed line) }\label{fig_compi}
\end{figure}

\begin{figure}
\centerline{\epsfxsize=2.8truein
\epsffile{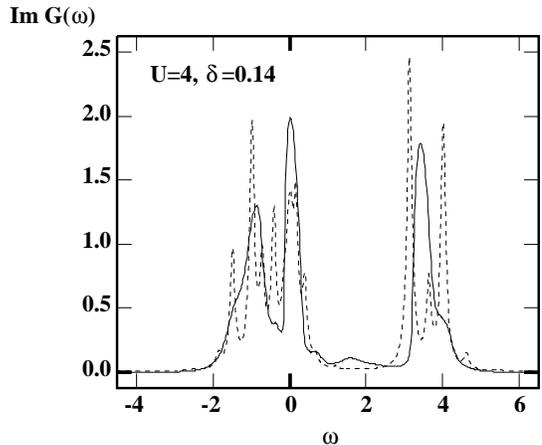}}
\caption{$\mbox{Im} G(\omega)$ at $T=0$
for $U=4D$ and hole doping $\delta=0.14$: iterative perturbation scheme
(full line) vs.~exact diagonalization (dashed line) }\label{fig_comp}
\end{figure}

To illustrate the   accuracy of our method 
we  compare  it with 
 results obtained using the exact  diagonalization
algorithm  of Caffarel and Krauth \cite{c4}.
Both methods are in close agreement when
used on the imaginary axis (see figure \ref{fig_compi}) . 
The real advantage of  
our perturbation scheme compared to the exact diagonalization
 is disclosed when we display the spectral functions obtained by
these two methods on the real axis (figure~\ref{fig_comp}).
It is clear that the exact diagonalization is doing its best in
producing the correct spectral distribution. But it is unable to give
a smooth density of states. Instead several sharp structures occur
as a consequence of treating only a finite number of orbitals in the
Anderson model.

\begin{figure}
\centerline{\epsfxsize=3.5truein
\epsffile{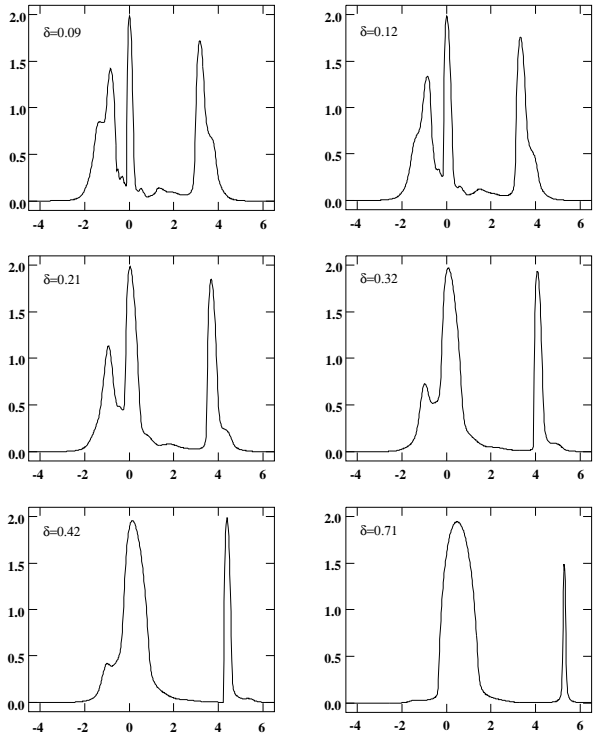}}
\caption{Evolution of the spectral function for $U=4\,D$ and $T=0$
with increasing  hole
doping $\delta$}\label{fig3}
\end{figure}

 Figure \ref{fig3} shows the evolution of the spectral
density 
 of the doped Mott insulator ($U=4\,D$\/) with increasing hole doping
$\delta$. 
The qualitative features are those expected from the spectra of the
single impurity \cite{cugeorges} and are in agreement with the quantum
Montecarlo calculations \cite{jarrell2}. 
For small doping there is a
clear resonance peak at the fermi level. As $\delta$ is increased, the peak
broadens and is shifted through the lower Hubbard band. At the same time
the weight of the upper band decreases.

The most striking feature of the evolution of the spectral density as
a function of doping is the finite shift of the Kondo resonance
from the insulating band edge 
as the doping goes to zero. It was demonstrated
analytically that this is a genuine property of the exact 
solution of the Hubbard model in infinite dimensions  
using the projective self-consistent method \cite{moeller2}
and is one of the most striking properties of the Hubbard model in large
dimensions.
This feature did not appear in   the earlier studies of Hubbard model in large
dimensions using Montecarlo techniques \cite{jarrell2} at higher 
temperatures, and is also  not easily seen in
exact diagonalization algorithms \cite{kkm}.

\section{Discussion and outlook} \label{su4}
In this paper we introduced a new perturbation scheme for the solution
of lattice models away from half filling. The basic idea is to
construct an expression for the self energy which interpolates between
correct limits. In the weak coupling limit
our approximate self energy is exact to order $U^2$,
and it is also exact in the 
atomic limit.
The proper low frequency behavior is ensured by the Friedel sum rule
(or, equivalently, the Luttinger theorem). This is important to obtain
the right low energy features in the spectral density. 
The overall distribution of the density of states on the other hand is
determined by the spectral moments, which are reproduced exactly
up to second order  by
satisfying the proper large frequency behavior.
In the light of these features it might not be too astonishing that
we obtain a good agreement with the exact diagonalization
method. 

Since the algorithm decribed here is accurate and very fast (a
typical run to solve the Hubbard model takes 60 seconds on a DEC alpha
station 200 4/233)
it has a wide range of applications.
Two examples that come to mind
 are the effects of disorder on the Hubbard model away
from half filling and the study of realistic models with orbital 
degeneracy. The latter is very important to make contact with 
realistic three dimensional transition metal oxides.

{\bf Acknowledgements:}
This work has been supported by the National Science Foundation.
DMR 92-24000

\end{document}